\documentclass{iau} 
\usepackage{graphicx}
\usepackage{natbib}

\title[Clarifying the HMXB population of the SMC] %% give here short title %%
{Clarifying the population of HMXBs in the Small Magellanic Cloud}

\author[Maravelias et al.]   %% give here short author list %%
{Grigoris Maravelias$^{1,2}$\thanks{contact: gmaravel@physics.uoc.gr}, Andreas Zezas$^{1,2,3}$, Vallia Antoniou$^{3}$, \\
Despina Hatzidimitriou$^{4,5}$, Frank Haberl$^{6}$}

\affiliation{
$^1$IESL, Foundation for Research and Technology-Hellas, Heraklion, Greece,\\  
$^2$Department of Physics, University of Crete, Heraklion, Greece, \\
$^3$Harvard-Smithsonian Center for Astrophysics, Cambridge, USA, \\
$^4$Department of Physics, National and Kapodistrian University of Athens, Zografou, Greece, \\
$^5$IAASARS, National Observatory of Athens, Athens, Greece \\
$^6$Max-Planck-Institut f\"{u}r extraterrestrische Physik, Garching, Germany
}

\pubyear{2015}
\volume{xxx}  %% insert here IAU Symposium No.
\jname{Title of your IAU Symposium}
\editors{A.C. Editor, B.D. Editor \& C.E. Editor, eds.}
\begin{document}

\maketitle

\begin{abstract}
Almost all confirmed optical counterparts of HMXBs in the SMC are OB stars with equatorial decretion disks (OBe). These sources emit strongly in Balmer lines and standout when imaged through narrow-band H$\alpha$ imaging. The lack of secure counterparts for a significant fraction of the HMXBs motivated us to search for more. Using the catalogs for OB/OBe stars \citep{Maravelias2017} and for HMXBs \citep{Haberl2016} we detect 70 optical counterparts (out of 104 covered by our survey). We provide the first identification of the optical counterpart to the source XTEJ0050-731. We verify that 17 previously uncertain optical counterparts are indeed the proper matches. Regarding 52 confirmed HMXBs (known optical counterparts with H$\alpha$ emission), we detect 39 as OBe and another 13 as OB stars. This allows a direct estimation of the fraction of active OBe stars in HMXBs that show H$\alpha$ emission at a given epoch to be at least $\sim75\%$ of their total HMXB population.

\end{abstract}

\firstsection % if your document starts with a section,
\section{Introduction}

The Small Magellanic Cloud (SMC) has been a major target for X-ray surveys for a number of reasons: (i) due to the complete coverage of the galaxy, (ii) our ability to detect sources down to non-outbursting X-ray luminosities $(L_X\sim10^{33}\,\textrm{erg s}^{-1})$, and (iii) its impressive large number of High-Mass X-ray Binaries \citep[HMXBs;][]{Haberl2016}. Thus, the SMC is a unique laboratory to examine HMXBs with a homogeneous and consistent approach. However, the X-ray properties alone cannot fully characterize the nature of each source. HMXBs consist of an early-type (OB) massive star and a compact object (neutron star or black hole), which accretes matter from the massive star either through strong stellar winds and/or Roche-lobe overflow in supergiant systems or through an equatorial decretion disk in, non-supergiant, OBe stars (Be/X-ray Binaries; BeXBs). Thus, to understand the nature of HMXBs in general we need to also study their optical counterparts. The OBe stars are massive stars that due to their decretion disks show Balmer lines in emission, of which H$\alpha$ is typically the most prominent (e.g., see the review by \citealt{Rivinius2013}). Although the SMC is close enough to resolve its stellar population, we still lack the identification of the optical counterparts or their nature for many HMXBs. Out of the $\sim120$ HMXBs (of which almost all are actually BeXBs; \citealt{Haberl2016}), the $\sim40\%$ is listed as candidate systems of this class of objects. By taking advantage of the fact that OBe stars display H$\alpha$ in emission (i.e., easily discernible in H$\alpha$ narrow-band images), we performed a wide area H$\alpha$ imaging survey of the SMC to reveal prime candidates for optical counterparts to HMXBs. 

\section{Survey and Catalog}

We used the Wide Field Imager (WFI@MPG/ESO 2.2m, La Silla, on 16/17 November, 2011) and the MOSAIC camera (@CTIO/Blanco 4m, Cerro Tololo, on 15/16 December, 2011) to observe 6 and 7 fields in the SMC, respectively, covering almost the entire main body of the galaxy. Each field was observed in the $R$ broad-band (the continuum) and H$\alpha$ narrow-band filters, in order to ensure the proper removal of the continuum and the calculation of H$\alpha$ excess (for details see \citealt{MaraveliasPhD,Maravelias2017}). The exposure time was selected to achieve a similar depth ($R\sim23$ mag) in both campaigns to cover all late B-type stars at the distance of the SMC. 

Using the locus of OB stars \citep{Antoniou2009} we selected the best OB candidate sources, for which we calculated their (H$\alpha-R$) index. From their corresponding distribution we estimate their mean (H$\alpha-R$) and their standard deviation $\sigma$. 
%Since the $R$ filter includes the H$\alpha$ region, the corresponding expected baseline for stars without any H$\alpha$ excess would be equal to (H$\alpha-R$)=0 mag. However, due to the differences between the two filters and the range of spectral types considered, this is not exactly equal to zero \citep{MaraveliasPhD}. To overcome this we define the baseline (H$\alpha-R$) value for non-H$\alpha$ excess stars individually for each field based on the mean ($\langle \textrm{H}\alpha-R\rangle$) and the standard deviation of the (H$\alpha-R$) distribution of all OB stars in each field. 
We consider as the most reliable H$\alpha$ emitting candidates the sources with: $(\textrm{H}\alpha-R) < \langle\textrm{H}\alpha-R\rangle - 5 \times \sigma$, and SNR$>5$ \citep{Maravelias2017}. The final catalog consists of $\sim8350$ H$\alpha$ emitting objects out of a parent population of $\sim77000$ OB stars.

\begin{figure}
\begin{center}
 \includegraphics[width=13cm,trim={5cm 0 3.5cm 0},clip]{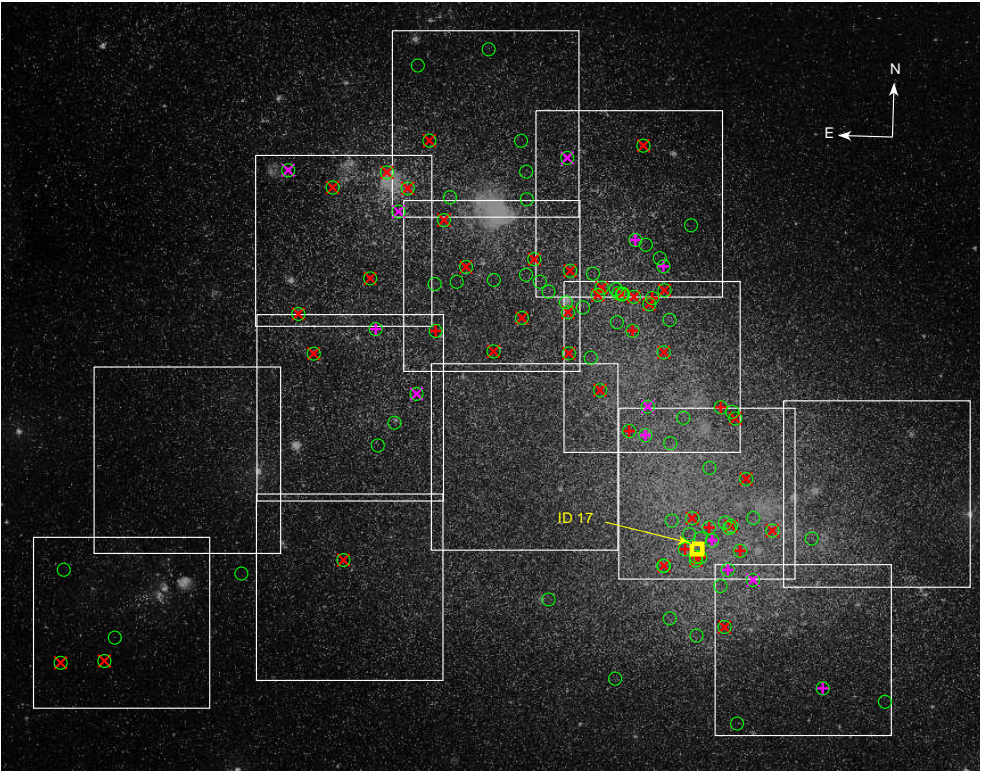} 
 \caption{Our observed fields (white boxes) overplotted on a DSS image of the Small Magellanic Cloud. The green circles correspond to HMXBs from the census of \cite{Haberl2016}, of which 104 lay within our fields and 70 have been identified. One of these is an H$\alpha$ emitting OB (yellow arrow/box) coinciding with the HMXB ID\#17 (\citealt{Haberl2016}; XTE J0050-731; SXP 16.6), an X-ray source without any optical identification previously. For 17 HMXBs we verify their, previously uncertain, optical counterparts to 7 OB stars (magenta crosses) and another 10 OB stars with H$\alpha$ emittion (red crosses). From 52 HMXBs with confirmed H$\alpha$ emitting counterparts we identify 39 sources (red X symbols) with H$\alpha$ excess, and 13 sources (magenta X symbols) are identified as OB stars (inactive OBe, i.e., without H$\alpha$ emission).}
   \label{fig}
\end{center}
\end{figure}

\section{Results}

In \cite{Maravelias2017} we have performed an initial analysis to derive the ratios of OB stars with emission (OBe) over their total parent OB population, as well as the HMXBs over the OBe population (OBe/OB$\sim13\%$ and HMXBs/OBe$\sim0.002-0.025\%$, respectively). In this work, we focus in a more detailed treatment of the HMXB population to identify the best optical counterparts. From the most recent census of HMXBs in the SMC by \cite{Haberl2016} we remove all rejected and unlikely HXMBs, along with 17 sources that lie outside our H$\alpha$ survey. Thus, we are left with 104 HMXBs, which we cross-correlate with our catalog to identify the counterparts of 70 HMXBs (see Fig. \ref{fig} and its caption for their representation).

We find that the optical counterpart of the source XTE J0050-731 or SXP16.6 (ID\#17 in the \citealt{Haberl2016} catalog) is an H$\alpha$ emitting OB star. This is the first identification of the optical counterpart of this source (not to be confused with RX J0051.8-7310),
%; \citealt{Galache2008})
and spectroscopy is needed to verify its nature. Moreover, we verify the, previously uncertain, optical counterparts for 17 sources to be OBe (10) or OB (7) stars. Last, we have detected 52 HMXBs, which have confirmed optical counterparts known to display H$\alpha$ in emission. Out of these sources we find 39 as H$\alpha$ emitters, while another 13 sources are identified only as OB stars (probably inactive Be stars due to their transient nature). Given these numbers, we estimate that at a given epoch $\sim75\%$ of HMXBs (BeXBs) are active. Although this is in general agreement with the fraction of active Be stars identified in open Galactic clusters ranging from 50-75\% \citep{Fabregat2003,McSwain2005,Granada2018}, we point out that it is actually a lower limit of the overall OBe population. This is because our selection criteria for OBe stars are conservative and this photometric approach is not sensitive to OBe stars with relatively faint H$\alpha$ emission. These results might differ from those based on the census of OB stars in clusters, as we examine a much larger OBe population covering both clusters and the field, in the environment of the SMC which has a much lower metallicity than the Galaxy. This consists a direct measurement of the actual fraction of active OBe stars in HMXBs (BeXBs) that show H$\alpha$ emission, i.e., an active decretion disk, at a given epoch. 

%\begin{acknowledgment}

GM acknowledges support by an IAU travel fund. GM and AZ acknowledge funding from the European Research Council under the European Union’s Seventh Framework Programme (FP/2007-2013)/ERC Grant Agreement n. 617001.

%\end{acknowledgment}

\end{document}